\documentclass
[aps,prl,twocolumn,floatfix,english,showpacs,10pt,superscriptaddress]{revtex4-2}%
\usepackage{graphicx}
\usepackage{booktabs}
\usepackage{amsmath}
\usepackage{physics}
\usepackage{amssymb}
\usepackage{colordvi}
\usepackage{verbatim}
\usepackage{xcolor}
\usepackage{mathrsfs}
\usepackage{epsfig}
\usepackage{lipsum}
\usepackage{amsfonts}
\usepackage{makecell}
\usepackage{esint}
\usepackage{bm}

\usepackage[most]{tcolorbox}

\usepackage[unicode=true, breaklinks=false, pdfborder={0 0 1}, backref=false,
colorlinks=true, linkcolor=blue, urlcolor=blue, citecolor=blue]{hyperref}%
\providecommand{\U}[1]{\protect\rule{.1in}{.1in}}

\providecommand{\U}[1]{\protect\rule{.1in}{.1in}}
\setcitestyle{numbers,square}

\begin{document}

\title{Universal Right-Hand Chirality of Evanescent Vector Fields}

\author{Yilin Qing}

\affiliation{School of Physics, Huazhong University of Science and Technology, Wuhan 430074, China}

\author{Ji Zou}
\email{ji.zou@kfupm.edu.sa}
\affiliation{Physics Department, King Fahd University of Petroleum and Minerals, 31261, Dhahran, Saudi Arabia}
\affiliation{Quantum Center, KFUPM, Dhahran, Saudi Arabia}

\author{Yi-Pu Wang}
\affiliation{Zhejiang Key Laboratory of Micro-Nano Quantum Chips and Quantum Control, School of Physics,  Zhejiang University, Hangzhou 310027, China}

\author{Gerrit E. W. Bauer}
\email{bauer.gerrit.ernst.wilhelm.d8@tohoku.ac.jp}
\affiliation{WPI-AIMR and IMR and CSIS, Tohoku University, Sendai 980-8577, Japan}

\author{Tao Yu}
\email{taoyuphy@hust.edu.cn}
\affiliation{School of Physics, Huazhong University of Science and Technology, Wuhan 430074, China}

\date{\today }

\begin{abstract}

A locking between propagation direction $\hat{\bf q}$, spin $\hat{\bf S}$, and surface or interface normal $\hat{\bf n}$---a phenomenon broadly termed \textit{chirality}---pervades evanescent or surface waves in optics, magnetism, plasmonics, and acoustics. 
Yet, it is not known whether the phenomenon is universal or bound to conditions. Here, we unveil that any \textit{source-free} vector field that propagates along  $\hat{\bf q}$ and is evanescent along $\hat{\bf n}$, the existence of a spin automatically enforces a rigid, right-handed locking among $\hat{\bf q}$, $\hat{\bf S}$, and $\hat{\bf n}$,  that is characterized by a chirality index $C_{\bf q}\equiv \hat{\bf n}\cdot(\hat{\bf S}\times \hat{\bf q})>0$. For an arbitrary propagation direction defined by $\hat{\bf q}\cdot \hat{\bf S}\equiv \eta\in [-1,1]$, i.e., not necessarily transverse spin, we derive a fundamental upper bound $C_{\bf q}\le \sqrt{1-\eta^2}$. The universal absence of left-handed evanescent (source-less) vector fields underscores the fundamental physical non-equivalence of mirror-image configurations---a principle echoed by parity violation, the natural excess of enantiomers in chiral molecules, and the fixed chirality of DNA.

\end{abstract}

\maketitle

\textit{Introduction}.---Chirality is a fundamental concept in nature, manifesting itself across a remarkable range of scales and disciplines, from parity violation in particle physics~\cite{lee1956question,wu1957experimental} and molecular chirality in chemistry~\cite{peluso2022recognition} to the double-helix structure of DNA in biology~\cite{watson1953molecular}. 
Wave phenomena are likewise ubiquitous. Chirality often emerges in waves as a dynamical handedness of field rotation or circular polarization that is locked to the propagation direction, most prominently in evanescent waves that decay exponentially as a function of distance from interfaces, boundaries, or guides~\cite{bliokh2014extraordinary,petersen2014chiral,electric_field_2,electric_field_3,bliokh2015spin,spin_momentum_locking_optics,chiral_nature_5,shi2021transverse}. Wave chirality causes directional wave-matter interactions~\cite{bliokh2014extraordinary,petersen2014chiral,electric_field_2,electric_field_3,bliokh2015spin,spin_momentum_locking_optics,chiral_nature_5,shi2021transverse,chiral_nature_2,chiral_nature_1,chiral_prl,magnetic_field_3,kuss2022chiral,zou2022bell,zou2024spatially,driessen2025robust,lininger2023chirality,Guozhi2026} and nonreciprocal transport~\cite{magnetic_field_1,electric_field_1,chiral_nature_3,chiral_nature_4,Zhang2019,yu2023chirality,Cheng2023,zou2024dissipative,nakata2024magnonic,zou2025emergent}, thereby enabling powerful functionalities in a wide range of physical platforms.

In magnetic systems, chirality is a hallmark of magnetostatic surface spin-wave modes, governing their unidirectional propagation~\cite{spin_wave2,spin_wave1,spin_wave3,yu2023chirality,yu2019chiral,cai2024spin,trevillian2024formation,xue2025directional}. In optics and plasmonics, the relation ${\bf q} \cdot {\bf S} =0$ locks the wave vector $\bf q$ of an evanescent wave to the transverse photon spin ${\bf S}$~\cite{spin_momentum_locking_optics,bliokh2014extraordinary,petersen2014chiral,bliokh2015spin,electric_field_1,electric_field_2,electric_field_3,VanMechelen2016}, attributed to the transversality and causality of Maxwell’s equations~\cite{VanMechelen2016,bliokh2014extraordinary}.  In acoustics and phononics, surface acoustic waves propagate with the same handedness \cite{SAW2,SAW1,shi2019observation,long2020realization,SAW3}, as do various phonon modes in metamaterials~\cite{yuan2021observation,long2020realization,yuan2021observation,Zhang2019,yang2023hybrid}. Moreover, the chirality in classical systems directly carries over into the quantum regime~\cite{chiral_nature_5,xue2025directional}. 
Despite their distinct physical origins and governing equations, these waves share a common geometric signature: propagation along an interface combined with exponential decay normal to it. In this Letter, we answer the intriguing question: under what minimal and general conditions does an evanescent vector field possess definite chirality?

\begin{figure}[htp!]
\centering
\includegraphics[width=1.0\linewidth]{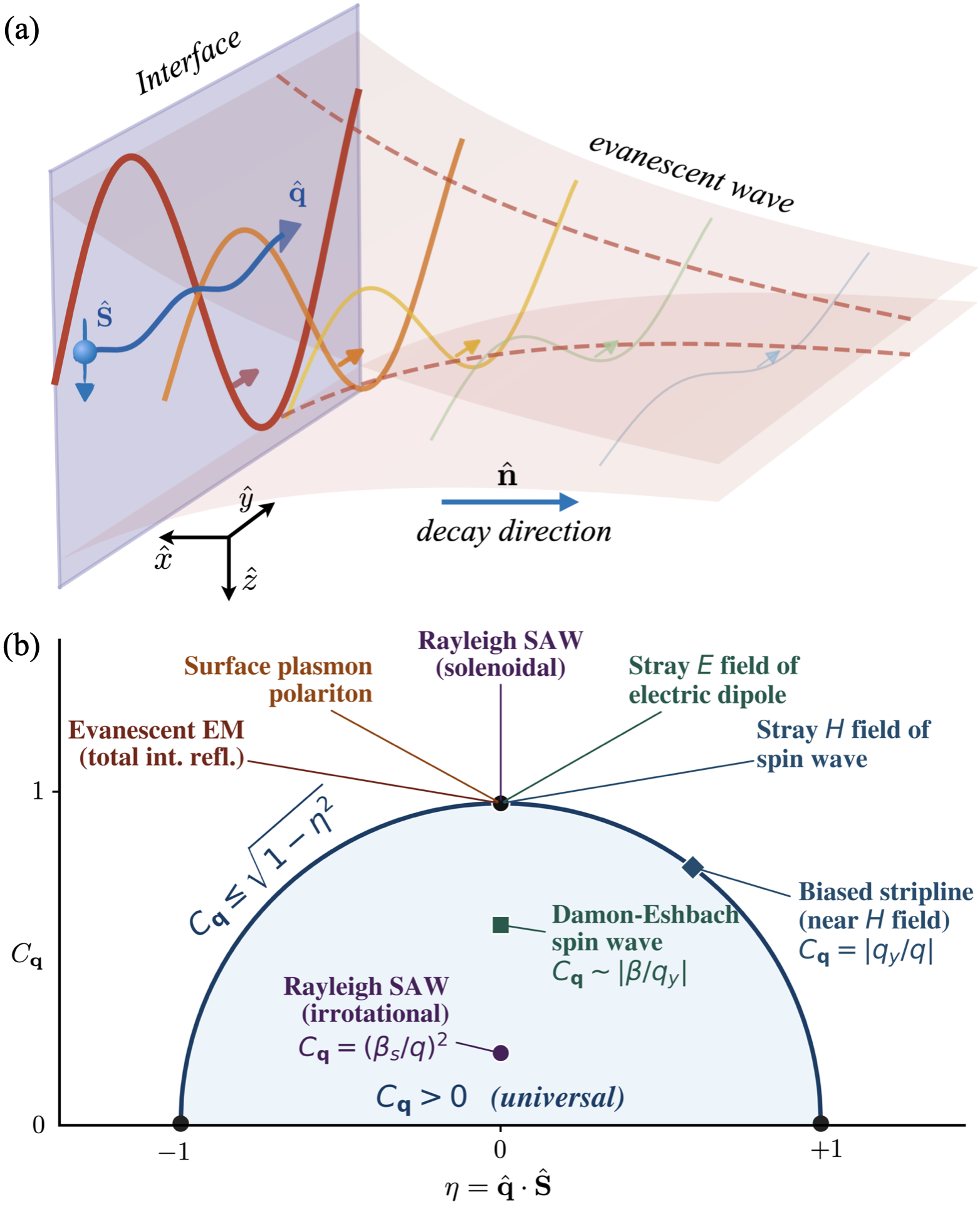}
\caption{(a) Artistic impression of a source-free evanescent vector field propagating along the interface in the $\hat{\bf y}$ direction, with exponentially decaying amplitude along the interface normal $-\hat{\mathbf x}$.
(b) Chirality index $C_{\mathbf{q}}$ vs. spin-orbit parameter $\eta$. 
The semicircle $C_{\mathbf{q}} = \sqrt{1-\eta^2}$ holds for elliptically polarized fields 
and is an upper bound for all propagation directions. Several prominent examples of evanescent waves are perfectly chiral with $C_{\bf q}=1$, but all of them fall into the upper semicircle with \(C_\textbf{q}>0\).}
\label{fig:1}
\end{figure}

We prove below that all \textit{spinful} and (nearly) \textit{source-free} evanescent vector field lock their propagation direction $\hat{\bf q}=\mathbf{q}/|{\bf q}|$ in the plane of an interface with surface normal $\hat{\bf n}$ (defined as the field decay direction) to the intrinsic angular momentum or spin direction $ \hat{\mathbf{S}}$
with a fixed right-handedness $C_\mathbf{q}=\hat{\bf n}\cdot(\hat{\bf S}\times \hat{\bf q})>0$, irrespective of the underlying medium. 
Figure~\ref{fig:1}(a) illustrates this universal right-handed rule. Evanescent electromagnetic waves, the solenoidal components of Rayleigh surface acoustic waves, stray electric fields of surface plasmon polaritons and single electric dipoles, and magnetic stray fields of circularly polarized spin-waves are fully chiral with $C_{\mathbf{q}} = 1$.  Further, we analyze the general case $0 \le C_{\mathbf{q}} \le 1$ in which the spin is not fully normal to the wave vector with a spin-orbit parameter $\eta\equiv \hat{\bf q}\cdot\hat{\bf S} \ne 0$ and $\eta\in [-1,1]$. Figure~\ref{fig:1}(b) shows the universal bounds $0 \le C_{\bf q}\le \sqrt{1-\eta^2}$ derived below, illustrated by the magnetic field emitted by a short stripline and Damon-Eshbach spin waves for wave vectors not precisely normal to the saturation magnetization. The absence of left-handed evanescent vector fields serves as another example of the non-equivalence of mirror-image processes in nature, similar to the parity violation of the weak nuclear interaction~\cite{lee1956question,wu1957experimental} and the chiral selection of molecules in biology~\cite{peluso2022recognition} including the fixed chirality of DNA~\cite{watson1953molecular}.

\textit{Universal chirality of evanescent vector fields}.---The physics of wave propagation is fundamentally governed by linear, constant-coefficient partial differential equations (PDEs), such as the Helmholtz, d'Alembert, or diffusion equations. Here, we consider a vector field of time-harmonic evanescent plane wave modes with frequency $\Omega$ and (real) in-plane wave vector \(\mathbf{q}=(0,q_y,q_z)\) in negative half space (\(x<0\)) with translational symmetry in the \((y,z)\)-plane
\begin{align}
\mathbf{V}(\mathbf{r},t)=\boldsymbol{\mathcal{V}} e^{i(q_y y+q_z z)+\beta_\mathbf{q} x}e^{-i\Omega t},
\label{vector_field}
\end{align}
which decays along the negative $x$-direction [Fig.~\ref{fig:1}(a)] on the scale $1/\beta_\mathbf{q} >0$. In evanescent electromagnetic waves $\beta_\mathbf{q}=\sqrt{q_y^2+q_z^2-(\Omega/c)^2}$ ~\cite{spin_momentum_locking_optics}, but in general $\beta_{\bf q}$ is a complicated function of ${\bf q}$ dictated by the specific PDE under consideration~\cite{SAW1,SAW2,SAW3,shi2019observation, spin_wave1,spin_wave2,spin_wave3}. The normalized field amplitudes $\boldsymbol{\mathcal{V}}=(1,\mathcal{V}_y e^{i\phi_y},\mathcal{V}_z e^{i\phi_z})$ contain four variables, viz. real-valued amplitudes $\mathcal{V}_y$ and $\mathcal{V}_z$ and the phases $\{\phi_y,\phi_z\} \in [0,2\pi)$. The vector field can be an electric field ${\bf E}$~\cite{electric_field_1,electric_field_2,chiral_nature_5,spin_momentum_locking_optics,petersen2014chiral,bliokh2015spin,electric_field_6,electric_field_5,electric_field_4}, a magnetic field ${\bf H}$~\cite{magnetic_field_1,magnetic_field_2,yu2019chiral,magnetic_field_3,magnetic_field_4}, a magnetization texture ${\bf M}$~\cite{spin_wave1,spin_wave2,spin_wave3,trevillian2024formation,zou2024dissipative}, a strain field ${\bf u}$~\cite{SAW1,SAW2,SAW3,shi2019observation,long2020realization,yang2023hybrid,yuan2021observation}, an electric-polarization texture ${\bf P}$~\cite{polarization_2,polarization_1,coherent_ferron}, or superconducting ${\bf d}$-vector~\cite{d_vector}.

The spin density 
\begin{equation}
    \mathbf{S}(\mathbf{r},t)=s\, \mathrm{Im}[\mathbf{V}^*(\mathbf{r},t)\times \mathbf{V}(\mathbf{r},t)]
    \label{spin}
\end{equation}
measures the degree of the field rotation, where $s$ accounts for the spin dimension. This definition agrees with the spin density of electric and magnetic fields $\{{\bf E}({\bf r},t),{\bf H}({\bf r},t)\}$ in vacuum~\cite{electric_field_1,electric_field_2,electric_field_3,electric_field_4,electric_field_5,electric_field_6,spin_momentum_locking_optics,petersen2014chiral,bliokh2015spin,magnetic_field_1,magnetic_field_2,yu2019chiral,magnetic_field_3,magnetic_field_4,chiral_nature_5}.  When $\mathbf{V}={\bf u}$~ is a strain field, \(\mathbf{S}\) is the phonon spin~\cite{SAW1,SAW2,SAW3,shi2019observation,long2020realization,yang2023hybrid,yuan2021observation,phonon_spin_1,phonon_spin_2,phonon_spin_3,chiral_phonon_1}. In an excited magnetization ${\bf M}({\bf r},t)-\mathbf{M}_0$,  Eq.~\eqref{spin} is the rotation direction of the magnetization textures (``magnon spin")~\cite{spin_wave1,spin_wave2,spin_wave3,trevillian2024formation,zou2024dissipative}, where the static magnetization \(\mathbf{M}_0\) may also have a nonzero spin. 
The spin density direction $\hat{\bf S}={\bf S}/|{\bf S}|$ of the field \eqref{vector_field} is a vector on the Bloch sphere with components that satisfy the relation
\begin{equation}
    q_z\mathcal{V}_y\sin\phi_y-q_y\mathcal{V}_z\sin\phi_z=\frac{\eta |{\bf q}||\bf{S}|}{2se^{2\beta x}}. \label{eq:general_spin_condition}
\end{equation}
The following derivation holds for \textit{any} source-free (``solenoidal") vector field with divergence  $\nabla \cdot {\bf V}(\mathbf{r},t)=0$ or
\begin{subequations}
\begin{align}
&q_y\mathcal{V}_y\cos\phi_y+q_z\mathcal{V}_z\cos\phi_z=0,  
\label{eq:divergence_free_real} \\
&q_y\mathcal{V}_y\sin\phi_y+q_z\mathcal{V}_z\sin\phi_z=\beta.  
\label{eq:divergence_free_imag}
\end{align}
\end{subequations}

Choosing a coordinate system such that $\hat{\bf n}=-\hat{\bf x}$, the chirality index
\begin{equation}
C_{\mathbf{q}}=-\hat{\mathbf{x}}\cdot(\hat{\mathbf{S}}\times\hat{\mathbf{q}})=-{\hat{\mathbf{x}}\cdot(\mathbf{S}\times\mathbf{q})}/({|\mathbf{S}||\mathbf{q}|}).
\end{equation}
The $x$-component $-(\mathbf{S}\times\mathbf{q})_x$ or
\begin{align}
S_z q_y-S_y q_z &= 2s(q_z\mathcal{V}_z\sin\phi_z+q_y\mathcal{V}_y\sin\phi_y)
e^{2\beta x}\nonumber\\
&=2s\beta e^{2\beta x}
\end{align}
governs the sign or handedness of the chirality. With modulus
\begin{align}
   & |\mathbf{S}|=2se^{2\beta x}\nonumber \\ 
   &\times \sqrt{(\mathcal{V}_y\sin\phi_y)^2+ 
    (\mathcal{V}_z\sin\phi_z)^2
    +\mathcal{V}_y^2 \mathcal{V}_z^2 \sin^2(\phi_y-\phi_z)},
    \label{eq:spin_magnitude_general}
\end{align} 
the chirality index
\begin{align}
C_\mathbf{q}&=\frac{\beta}{|{\bf q}|\sqrt{(\mathcal{V}_y\sin\phi_y)^2+(\mathcal{V}_z\sin\phi_z)^2+\mathcal{V}_y^2 \mathcal{V}_z^2 \sin^2(\phi_y-\phi_z)}}\nonumber\\
&>0.
\label{general_chirality_index}
\end{align} 
All source-free evanescence/surface modes, therefore, automatically lock their decay, spin, and propagation directions into a right-hand rule. 
The degree of the chirality is a complicated function of several parameters $\{q_y,q_z,\eta,{\cal V}_y,{\cal V}_z,\phi_y,\phi_z,x\}$, however.

When $\phi_{y/z}/\pi$ are integers, $|\mathbf{S}|=C_\mathbf{q}=0$. In the special cases that $\phi_z=\phi_y$ or $\phi_z-\phi_y=\pm \pi$ the source-free conditions \eqref{eq:divergence_free_real} and \eqref{eq:divergence_free_imag} demand that $\phi_z$ and $\phi_y$ can take only specific values: when they are equal  $\phi_z=\phi_y={\pi}/{2}$ or $\phi_z=\phi_y={3\pi}/{2}$; otherwise $\{\phi_y,\phi_z\}=\{{\pi}/{2},{3\pi}/{2}\}$ or $\{\phi_y,\phi_z\}=\{{3\pi}/{2},{\pi}/{2}\}$. The chirality indices then simplify to
\begin{equation}
    C_\mathbf{q}=\sqrt{1-\eta^2}.
    \label{simpleform}
\end{equation}
It becomes unity for \(\eta=0\), i.e., transverse spin: The three vectors $\{\hat{\bf n},\hat{\bf S}, \hat{\bf q}\}$ are then normal to each other and are locked by the right-hand rule.

Otherwise, we multiply Eq.~\eqref{eq:divergence_free_real} by $\sin\phi_y\sin\phi_z$ and Eq.~\eqref{eq:divergence_free_imag} by $\cos\phi_y\cos\phi_z$,  and subsequently divide by $\sin(\phi_z - \phi_y) \ne 0$. Substituting the amplitudes
\begin{equation}
    \mathcal{V}_y = - \frac{\beta \cos \phi_z}{q_y \sin(\phi_z - \phi_y)}, \nonumber\quad\mathcal{V}_z = \frac{\beta \cos \phi_y}{q_z \sin(\phi_z - \phi_y)},
    \label{solutions}
\end{equation}
into Eq.~\eqref{eq:general_spin_condition} yields the constraint 
\begin{align}
\left(\frac{q_z}{q_y}\right)^2+\frac{\eta|{\bf q}|\sqrt{q_z^2+(q_y\Phi)^2+(\beta\cot\phi_y)^2}}{{\operatorname{sgn}(\Phi-1)|q_zq_y|}} \left( \frac{q_z}{q_y}\right)+\Phi=0,
\label{quadratic_equation}
\end{align}
where $\Phi(\phi_y,\phi_z)={\cot\phi_y}/{\cot\phi_z}$. When $|\mathbf{q}|\ne 0$,  $(q_z/q_y)^2$ obeys the quartic equation 
\begin{align}
&\left({\beta^2\eta^2\cot^2\phi_y}/{q^2}+\eta^2-1\right)\left({q_z}/{q_y}\right)^4\nonumber\\
&+\left({2\beta^2\eta^2\cot^2\phi_y}/{q^2}+\eta^2\Phi^2+\eta^2-2\Phi\right)\left({q_z}/{q_y}\right)^2\nonumber\\
&+\left({\beta^2\eta^2\cot^2\phi_y}/{q^2}+\eta^2\Phi^2-\Phi^2\right)=0,
\label{uequation}
\end{align}
which is solved by
\begin{align}
\left(\frac{q_z}{q_y}\right)^2_{\pm} = \frac{-(\frac{2\beta^2\eta^2\cot^2\phi_y}{q^2}+\eta^2\Phi^2+\eta^2-2\Phi) \pm  \sqrt{\Delta}}{2\left(\frac{\beta^2\eta^2\cot^2\phi_y}{q^2}+\eta^2-1\right)},
\label{two_solutions}
\end{align}
where $\Delta = (\Phi-1)^2 \eta^2\left[{4\beta^2\cot^2\phi_y}/{q^2}+\eta^2(\Phi+1)^2-4\Phi \right]\ge 0$. When $\eta\ne 0$, the chirality index simplifies to
\begin{equation}
C_{\mathbf{q}}(\eta\ne 0)=\left| \frac{({q_z}/{q_y})(\Phi-1)\eta}{({q_z}/{q_y})^2+\Phi}\right|,
\label{simpleindex}
\end{equation}
in which ${q_z}/{q_y}$ solves Eq.~\eqref{uequation}. Otherwise,
\begin{align}
    C_\mathbf{q}(\eta=0)=\frac{1}{\sqrt{1-({\beta^2}/{|\mathbf{q}|^2})\cot\phi_y\cot\phi_z}}.
\end{align}
In the special case $q_{y}=0$ (or $q_z=0$),  $\phi_{z} = \{\pi/2,3\pi/2\}$ (or $\phi_{y} = \{\pi/2,3\pi/2\}$) and $\mathcal{V}_{z} = \pm \beta/q$ (or $\mathcal{V}_{y} = \pm \beta/q$), such that the chirality index
\begin{equation}
  C_\mathbf{q}(q_{y}=0~{\rm or}~q_z=0)=  \sqrt{ 1 - \eta^2 \left( 1 + \frac{\beta^2}{{|\mathbf{q}|}^2}\cot^2\phi_y \right) }.
\end{equation}
When $\eta=0$, we find $C_{\bf q}(q_y=0)$ to be unity with ${\bf q}=q_z\hat{\bf z}$, since the spin ${\bf S}\parallel \hat{\bf y}$ is transverse.

The attenuation factor $\beta({\bf q})$ importantly affects the chirality. For the electromagnetic waves in  vacuum, $\beta\sim |{\bf q}|$~\cite{electric_field_1,electric_field_2,electric_field_3,electric_field_4,electric_field_5,electric_field_6,spin_momentum_locking_optics,petersen2014chiral,bliokh2015spin,magnetic_field_1,magnetic_field_2,yu2019chiral,magnetic_field_3,magnetic_field_4,chiral_nature_5}, but it may be smaller than that~\cite{spin_wave1,spin_wave2,trevillian2024formation,zou2024dissipative,SAW1,SAW2,SAW3,shi2019observation,long2020realization,yang2023hybrid,yuan2021observation}. When $\beta\ll |{\bf q}|$, disregarding terms $\propto (\beta/q)^2$ in Eq.~\eqref{uequation} leads to
\begin{align}
   \left(\frac{q_z}{q_y}\right)^2+\Phi=\pm\frac{\eta(\Phi-1)}{\sqrt{1-\eta^2}}\frac{q_z}{q_y},
\end{align}
in which  $\Phi\in\left(-\infty,\frac{1-\sqrt{1-\eta^2}}{1+\sqrt{1-\eta^2}}\right)\cup\left(\frac{1+\sqrt{1-\eta^2}}{1-\sqrt{1-\eta^2}},+\infty\right)$ split into two sheets.
Substitution of it into Eq.~\eqref{simpleindex} yields the simple expression  $C_\mathbf{q}(\beta\ll |{\bf q}|)=\sqrt{1-\eta^2}$.

\begin{figure}[htp!]
    \centering
    \includegraphics[width=\linewidth, clip, trim=0.6cm 0.25cm 0.4cm 0.1cm]{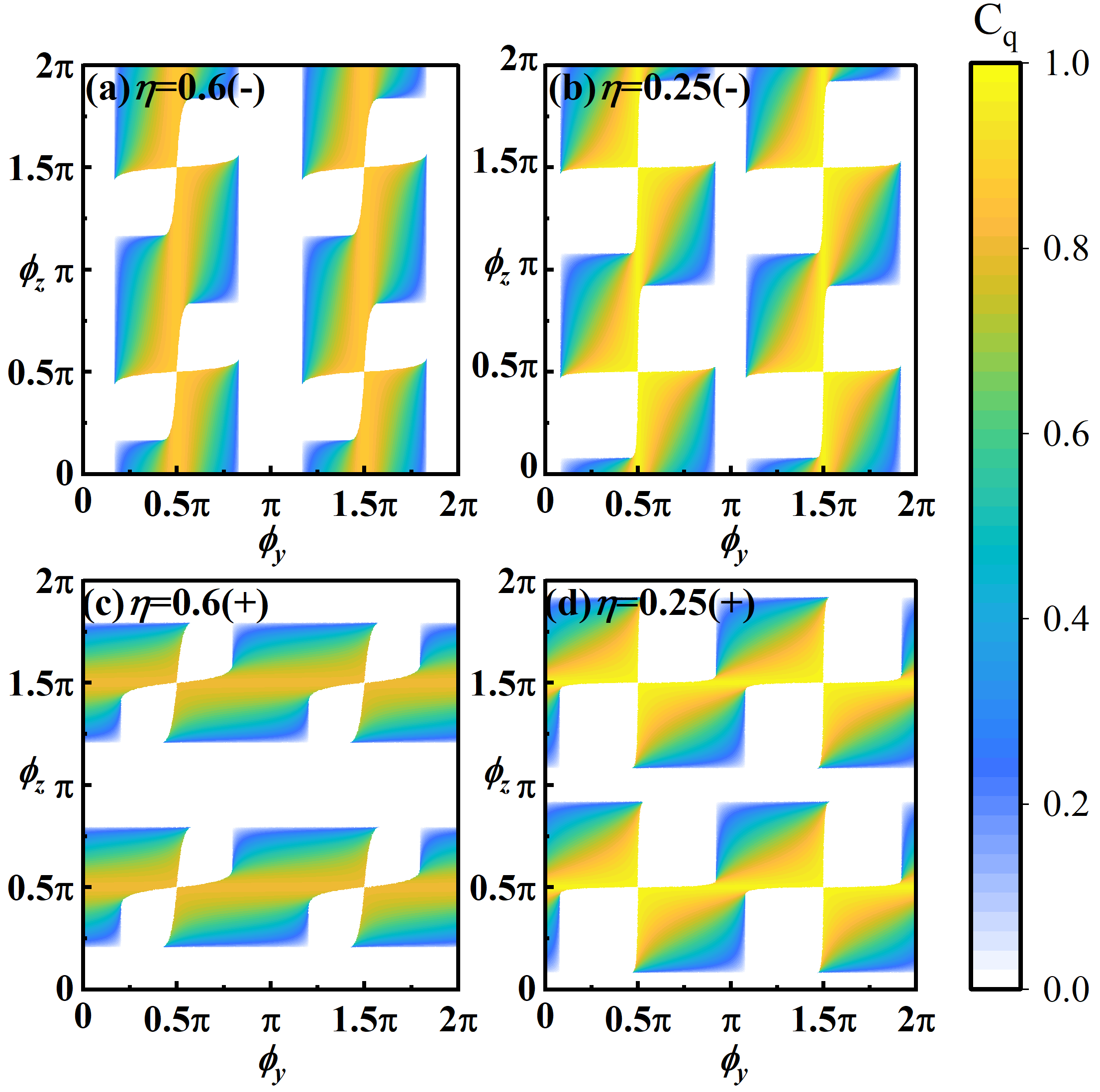}
    \caption{Universal dependence of the chirality index $C_{\bf q}$ on the phases $\{\phi_y,\phi_z\}$ for different spin-orbit parameter $\eta$.  ``$-$" in (a,b) and  ``+" in (c,d) indicate the two solutions $(q_z/q_y)_{+}$ and $(q_z/q_y)_{-}$ in Eq.~\eqref{two_solutions}. In the calculation, we take ${\beta}=|\mathbf{q}|$.}
    \label{fig:chirality_index}
\end{figure}

\begin{table*}[htbp]
\centering
\caption{Chirality of prototypical evanescent waves in nature. A time-harmonic dependence of $e^{-i\Omega t}$ with frequency $\Omega$ is implicitly assumed for all dynamic fields.}
\label{tab:chiral_waves}
\renewcommand{\arraystretch}{1.5}
\begin{tabular}{|p{3.7cm}|p{7cm}|p{3.8cm}|p{2.6cm}|}
\hline
\hline
\textbf{Wave type} & \textbf{Field components}  & \textbf{Normalized amplitudes} & \textbf{Chirality index} $C_\mathbf{q}$ \\
\hline
% Row 1: Circular TIR
Evanescent electromagnetic waves at total internal reflection  & 
$\begin{aligned}
E_{x} &\propto -i \frac{q_y}{\beta} e^{\beta x} e^{i q_y y}\\
  E_{y} &\propto e^{\beta x} e^{i q_y y}
\end{aligned}$ & 
$\left(1,i\frac{\beta}{q_y},0\right)$ & $1$\\
\hline
% Row 2: Surface Plasmon Polariton
Electric fields emitted by surface plasmon polaritons  & 
$\begin{aligned}
E_x &\propto q_y e^{\beta x} e^{iq_y y}\\
E_y &\propto i\beta e^{\beta x} e^{iq_y y}
\end{aligned}$ & 
$\left(1,i\frac{\beta}{q_y},0\right)$ &$1$ \\
\hline
% Row 3: electric dipoles
Stray electric fields of dynamical electric dipoles & 
$\begin{aligned}
E_x &\propto iq_ye^{|q_y|x}e^{iq_y y}\\
E_y &\propto-|q_y|e^{|q_y|x}e^{iq_y y}
\end{aligned}$ & 
$\left(1,i\text{sgn}(q_y),0\right)$ &$1$ \\
\hline
% Row 4: DE modes
Damon-Eshbach surface spin waves (the film occupies $[-d,0]$; $a_{\bf q}\sim -1$ and $b_{\bf q}\lesssim -1$) & 
$\mathbf{M}(q_y<0)|_{x>0}\propto 
\begin{pmatrix}
   -a_{\bf q} \beta + b_{\bf q} q_y\\
   i\left(-b_{\bf q} \beta + a_{\bf q} q_y\right)\\0
\end{pmatrix}e^{-\beta x}e^{i q_yy}$ & 
$ \left( 1, \ i \frac{a_{\bf q} q_y - b_{\bf q} \beta}{b_{\bf q} q_y - a_{\bf q} \beta}, \ 0 \right)$ & $\left| \frac{\beta (b_{\bf q} q_y - a_{\bf q} \beta)}{q_y (a_{\bf q} q_y - b_{\bf q} \beta)} \right|\newline\sim \left|\frac{\beta}{q_y}\right|$ \\
\cline{2-4}
&  
$\mathbf{M}(q_y>0)|_{x<-d}\propto
\begin{pmatrix}
     a_{\bf q} \beta + b_{\bf q} q_y\\i(b_{\bf q} \beta + a_{\bf q} q_y)\\0
\end{pmatrix}e^{\beta (x+d)}e^{i q_yy}$ & $ \left( 1, \ i \frac{b_{\bf q} \beta + a_{\bf q} q_y}{a_{\bf q} \beta + b_{\bf q} q_y}, \ 0 \right)$& $ \left| \frac{\beta (a_{\bf q} \beta + b_{\bf q} q_y)}{q_y (b_{\bf q} \beta + a_{\bf q} q_y)} \right|\newline\sim \left|\frac{\beta}{q_y}\right|$\\ 
\hline
% Row 5: Rayleigh SAW
Rayleigh surface acoustic waves ($\beta_s$ and $\beta_v$ are decay rates for irrotational and solenoidal components) & 
$\begin{aligned}
\mathbf{u}_{\text{irrotational}}&\propto \begin{pmatrix} \beta_s \\0\\ iq \end{pmatrix} e^{\beta_s x} e^{iqz}
\end{aligned}$ & 
$\left( 1,\ 0,\ i\frac{q}{\beta_s} \right)$ &$ \left(\frac{\beta_s}{q}\right)^2$ \\
\cline{2-4}
& 
$\begin{aligned}
\mathbf{u}_{\text{solenoidal}}&\propto \begin{pmatrix} - \frac{2q^2\beta_s}{q^2 + \beta_v^2} \\0\\ -i \frac{2q\beta_s\beta_v}{q^2 + \beta_v^2} \end{pmatrix} e^{\beta_v x} e^{iqz}
\end{aligned}$ & $ \left( 1,\ 0,\ i\frac{\beta_v}{q} \right)$&$1$ \\
\hline
% Row 6: Stripline Field
Near magnetic field of biased stripline ($\delta$ is the thickness of the stripline) & 
$\begin{aligned}
     H_x &\propto -2ie^{\beta x}\frac{e^{-\beta \delta}-1}{\beta^2}  e^{-i(q_yy+q_z z)}  \\
    H_y&\propto 2e^{\beta x}\frac{e^{-\beta \delta}-1}{\beta q_y}e^{-i(q_yy+q_z z)} 
\end{aligned}$ & 
$\left(1,i\frac{\beta}{q_y},0\right)$ & $\left|\frac{q_y}{\sqrt{q_y^2+q_z^2}}\right|$\\
\hline
% Row 7: Dipolar stray magnetic fields
Stray magnetic fields of spin waves (the magnetic film occupies $[0,\delta]$)& 
$\begin{aligned}
\begin{pmatrix} 
H_x \\ 
H_y 
\end{pmatrix}_{x>\delta} &\propto \frac{|q_y| + q_y}{2} \begin{pmatrix} 1 \\ -i \end{pmatrix} e^{-|q_y|(x-\delta)} e^{iq_y y}
\end{aligned}$ &$\left(1, -i ,0\right)$ & 1\\
\cline{2-4}
& 
$\begin{aligned}
\begin{pmatrix} 
H_x \\ 
H_y 
\end{pmatrix}_{x<0} &\propto \frac{|q_y| - q_y}{2} \begin{pmatrix} 1 \\ -i \end{pmatrix} e^{|q_y|x} e^{iq_y y}
\end{aligned}$ &$(1, -i ,0)$ & 1\\
\hline
\end{tabular}
\end{table*}

Although the chirality index $C_{\bf q}$ is parameter dependent, we find $\sqrt{1-\eta^2}$ is its upper bound. From Eq.~\eqref{uequation}, 
\[
\frac{\beta^2\cot^2\phi_y}{q^2}\eta^2 \frac{ \left[ \left(\frac{q_z}{q_y}\right)^2 + 1 \right]^2 }{ \left[ \left(\frac{q_z}{q_y}\right)^2 + \Phi \right]^2 } = (1-\eta^2) - \frac{ \eta^2(\Phi-1)^2\left(\frac{q_z}{q_y}\right)^2 }{ \left[ \left(\frac{q_z}{q_y}\right)^2 + \Phi \right]^2 }.
\]
Noting the second term at the right-hand side of this equation is exactly $C_{\mathbf{q}}^2$, we find 
\begin{align}
C_{\mathbf{q}}^2 = (1-\eta^2) - \frac{\beta^2\cot^2\phi_y}{q^2}\eta^2 \left( \frac{ \left({q_z}/{q_y}\right)^2 + 1 }{ \left({q_z}/{q_y}\right)^2 + \Phi } \right)^2.
\end{align}
Since 
\[
\frac{\beta^2\cot^2\phi_y}{q^2}\eta^2 \left( \frac{ \left({q_z}/{q_y}\right)^2 + 1 }{ \left({q_z}/{q_y}\right)^2 + \Phi } \right)^2 \ge 0,
\]
the chirality index obeys a universal upper bound:
\[
C_{\mathbf{q}} \le \sqrt{1-\eta^2}.
\]

According to the above analysis of different situations, the chirality of evanescent vector fields is governed by only three parameters $\{\phi_y,\phi_z,\eta\}$, irrespective of the physical medium.

 Figure~\ref{fig:chirality_index} illustrates the dependence of the chirality index $C_{\bf q}$ on the phase factors $\{\phi_y,\phi_z\}$. It shows universal features that only depend on $\eta$: i),  $\eta$ does not significantly affect the shape of the islands with $C_{\bf q}\ne 0$; ii) The chirality index is large  when one of $\{\phi_y,\phi_z\}$ is around $\{\pi/2,3\pi/2\}$. 
 
Table \ref{tab:chiral_waves} illustrates typical evanescent waves in nature, all exhibiting the universal chirality, delegating the derivations to the Supplemental Material~\cite{supplement}.

Validating the theoretical prediction of universal chirality by experiments on photonic, phononic, or magnonic platforms and metamaterials enables functionalities rooted in the spin-momentum and spin-surface locking.

\textit{Breaking the chirality by sources}.---A source that breaks the chirality of evanescent waves is a ``charge" \(\rho\) that oscillates with frequency $\Omega$ and wave vector ${\bf q}=q_y\hat{\bf y}+q_z\hat{\bf z}$ in the Poisson equation
\[
\nabla\cdot \mathbf{V}(\mathbf{r},t)=\rho e^{i(q_yy+q_zz)+\beta x}e^{-i\Omega t},
\]  
in which \(\rho=a+ib\in \mathbb{C}\) is a complex number with \(a,b\in \mathbb{R}\), or
\begin{align}
&q_y\mathcal{V}_y\sin\phi_y+q_z\mathcal{V}_z\sin\phi_z = \beta - a, \nonumber\\
&q_y\mathcal{V}_y\cos\phi_y+q_z\mathcal{V}_z\cos\phi_z = b,
\label{eq:divergent_imag}
\end{align}
which leads to the chirality index
\[
C_{\bf q}\propto (-\mathbf{S}\times\mathbf{q})_x = S_zq_y - S_yq_z = 2s(\beta-a) e^{2\beta x}
\]  
that depends on \(\beta-a\). A sufficiently strong source \(a\) can change its sign and break the chirality. A typical example is exchange-surface spin waves in antiferromagnets~\cite{Sun_exchange} that lack chirality because the magnetization and N\'eel vector fields violate the source-free condition (see Supplemental Sec.~VIII~\cite{supplement}).

\textit{Conclusion and discussion}.---In summary, we have proven that all source-free evanescent vector fields governed by a single exponential decay obey an intrinsic right-handed chirality,
$C_q>0$, bounded by $\sqrt{1-\eta^2}$. The locking of wave vector, spin, and surface normal imposes strong constraints on wave propagation---manifesting as a non-relativistic spin–orbit interaction. Broken time-reversal symmetry breaks the spin degeneracy, forcing surface waves to be unidirectional as in Damon–Eshbach surface spin waves and dipolar stray fields of spin waves. In time-reversal symmetric systems, the spin direction is arbitrary, allowing propagation in both directions on a given surface, yet the spin remains tied to the wave vector via the right-handed rule. This accounts for the properties of waves such as electric fields in surface plasmons, surface acoustic waves, stray electric fields of electric dipoles, and near magnetic fields in microwave striplines. Beyond unifying known phenomena, this theorem provides a diagnostic principle: observation of $C_{\bf q}\le 0$ would directly signal the presence of bulk sources, higher multipoles, or sublattice-scale structure beyond the continuum description.

\begin{acknowledgments}
This work is financially supported by the National Key Research and Development Program of China under Grant No.~2023YFA1406600 and the National Natural Science Foundation of China under Grant No.~12374109. J.Z. acknowledges the Quantum Center for the support received under No.~INQC2600. Y.P.W. acknowledges support from the National Key Research and Development Program of China (No.~2023YFA1406703 and 2022YFA1405200) and the Zhejiang Provincial Natural Science Foundation of China (No.~LR26A040001). G.E.W.B. acknowledges support from JSPS KAKENHI Grants No.~22H04965 and No.~24H02231 and Kei Yamamoto for pointing out the non-chirality of exchange surface waves in antiferromagnets.
\end{acknowledgments}

\bibliography{bib}

\end{document}